\documentclass[%
 aip,
 amsmath,amssymb,
reprint, onecolumn %%%%% Comment onecolumn to override one column format %%%%%%%
]{revtex4-1}

\usepackage{graphicx}% Include figure files
\usepackage{dcolumn}% Align table columns on decimal point
\usepackage{bm}% bold math
\usepackage[utf8]{inputenc}
\usepackage[T1]{fontenc}
\usepackage{mathptmx}
\usepackage{etoolbox}
\usepackage{subcaption}
\usepackage{multirow}
\usepackage{array}

\makeatletter
\def\@email#1#2{%
 \endgroup
 \patchcmd{\titleblock@produce}
  {\frontmatter@RRAPformat}
  {\frontmatter@RRAPformat{\produce@RRAP{*#1\href{mailto:#2}{#2}}}\frontmatter@RRAPformat}
  {}{}
}%
\makeatother
\begin{document}

\preprint{AIP/123-QED}

%\title[Sample title]{Sample Title:\\with Forced Linebreak}
\title{Surface and sub-surface modifications of copper electrodes exposed to electric high-field conditioning at cryogenic temperatures.}

\author{Marek Jacewicz}%
 \email{marek.jacewicz@physics.uu.se}
\affiliation{ 
Department of Physics and Astronomy, Uppsala University, Uppsala, Sweden%
}
\author{Iaroslava Profatilova}
\affiliation{ 
Department of Physics and Astronomy, Uppsala University, Uppsala, Sweden%
}
\author{Yinon Ashkenazy}
\affiliation{ 
Recah Institute of Physics, The Hebrew University of Jerusalem, 9190401 Jerusalem, Israel%
}
\author{Sergio Calatroni}
\affiliation{ 
CERN, European Organization for Nuclear Research, 1211 Geneva, Switzerland%
}
\author{Inna Popov}
\affiliation{ 
Recah Institute of Physics, The Hebrew University of Jerusalem, 9190401 Jerusalem, Israel%
}
\author{Piotr Szaniawski}
\affiliation{ 
Department of Physics and Astronomy, Uppsala University, Uppsala, Sweden%
}
\author{Walter Wuensch}
\affiliation{ 
CERN, European Organization for Nuclear Research, 1211 Geneva, Switzerland%
}

\date{\today}% It is always \today, today,

\begin{abstract}

It has been stipulated that the conditioning of metal surfaces under a high electric field is dominated by material hardening. To test this, we subjected three pairs of copper electrodes to high-voltage conditioning at distinct temperatures (300~K, 30~K, and 10~K) until each reached its saturation field.
The sets conditioned at colder temperatures showed a significant increase (36 \%) in the field holding capability with respect to the room temperature sample. The samples were then investigated with high-resolution microscopy, characterizing the breakdown (BD) spots on the anode and cathode according to their morphology, and with STEM, analyzing the changes in the subsurface regions. A self-shielding mechanism is offered to explain the observation of a central protected region within the anode spot. Unusual BD features were found on the cold-conditioned cathode surfaces, with very shallow craters of a star-like shape. The number of atypical spots increased with decreasing temperatures, reaching 26 and 53 percent of the total number of spots at 30~K and 10~K, respectively, and its form was suggested to be related to the different dynamics of the evolution of the spots due to thermal diffusivity variations. Sub-surface analysis showed clear structural changes in high-field areas, with stronger effects in the cold-conditioned sample. These suggested considerable sub-yield plastic activity during conditioning, leading to the initial observation of the formation of dislocation-denuded zones close to the surface. Thus, while conditioning is a result of sub-surface plastic evolution, it is not necessarily a result of surface hardening.

\end{abstract}

\maketitle

\section{Introduction}

High electrical fields in ultra-high vacuum environments (UHV) are needed in many areas of research and for many technologies, such as high-energy accelerators, high-vacuum interrupters, fusion research, electron and X-ray sources and space applications ~\cite{handbook_vacuum_arcs,clic_report}. The upper limit to which electrodes can hold an electric field is a major consideration in the design and operation of various systems.
It has been observed that when high electrical field pulses are applied to the surface repeatedly, the field-holding level increases~\cite{descoeudres_dc_2009, Degiovanni2016}. An explanation for this conditioning process is that the sharp features on the electrode surface are destroyed during the breakdown (BD), and so higher fields can be applied to the surface without further breakdowns. But experimental data show that this is not a complete explanation. Firstly, even very smooth surfaces, if unconditioned, will eventually experience breakdowns at a given field, but will accept higher fields after conditioning, with the much rougher surface covered by the breakdown craters. 
Moreover, the maximum field that a metal surface can tolerate is well correlated with its crystal structure~\cite{material_comparison, descoeudres_dc_2009}. 
Furthermore, in~\cite{Degiovanni2016}, it has been shown that the conditioning effect correlates with the number of pulses applied, and not with the number of breakdowns, suggesting that the conditioning is an effect of pulses and not of breakdowns.
Breakdowns are, at best, a way of measuring the conditioning state of a surface.

A more complex picture has emerged, in which there are extrinsic and intrinsic mechanisms responsible for the breakdown generation~\cite{Saressalo:2021kwm, BagchiPhysRevAccelBeams, wang_ab_2022, korsback2020}. 
The former are the aforementioned irregularities or contamination of the electrode surface. The latter mechanism is related to the material properties and alters the breakdown rate even when the effect of extrinsic processes was exhausted by cleaning of the surface via initial "processing" by plasma from discharges. The research into the mechanisms behind intrinsic breakdown generation has been an active field of study for many years and has gained renewed interest over the last decade. 
In recent years, theoretical models concentrated on trying to link pre-breakdown plastic activity at a metallic surface exposed to a high field to the formation of a localized specific feature, which will nucleate the ensuing breakdown. Various options for such local evolution were studied, including local dislocation activity ~\cite{pohjonen_dislocation_2011, pang_dislocation_2014}. 
These stress-driven processes have characteristic thermal activation energies, resulting in an exponential temperature dependence of breakdown rates, regardless of the specific nature of the defect reaction~\cite{Nordlund:2012zz, Zadin_2018}.

Recently, an effective kinetic mean field model, which describes the evolution of a self-interacting dislocation population, led to a similar dependence
even when self-interactions between dislocations and various obstacles are taken into account~\cite{Engelberg2018PRL}.
Indications to the validity of this description were found in measurements of pre-breakdown dark current fluctuations, which are consistent with the dynamics depicted in the model. The measured distribution fits well with the model prediction~\cite{Engelberg:2020PRAB, Paszkiewicz:2020sce}.
However, due to experimental difficulties, as of now there is no direct observation linking specific dislocation activity to breakdowns, and we are left with non-direct measurements~\cite{Ashkenazy2022}. 

Since most kinetic models include a thermally activated process, observing the temperature dependence of the breakdown processes is expected to offer a significant change in the breakdown properties, which can be measured and used to constrain various theoretical models~\cite{Engelberg:2020PRAB}.
Studies have shown that the use of cold temperatures will affect conditioning by largely increasing the maximal attainable fields on the surface. These changes are predicted by the kinetic models described above. However, it should be noted that changes in material properties such as hardness, thermal conduction, and thermal expansion coefficients may lead to such changes even if these models are not held.
The role of cryogenic temperatures on the BD rate has been verified experimentally both in radio-frequency (RF) and in DC studies~\cite{nasr_experimental_2021, Jacewicz2020}.

In this paper, we present a study combining the investigation of the intrinsic and extrinsic mechanisms by observing surface modification due to conditioning in cryogenic temperatures under very high fields. 
 
Conditioning was previously shown to depend on the hardness of Cu electrodes and was suggested to be related to microstructural evolution caused by the stress associated with the applied electric field~\cite{korsback2020}. Thus, we compared samples conditioned at room and cryogenic temperatures using various microscopy techniques to characterize the sub-surface and surface features found in the samples.

\section{Experimental}
\label{sec:method}

\subsection{Electrodes}

The results reported here are from measurements on three OFE copper samples. 
Electrodes were produced from copper UNS~C10100~Grade~1, which was proposed in the past as a candidate material 
for high gradient RF cavities~\cite{Aicheler2012AReport}. The electrodes were produced using fly-cut followed by diamond turning, achieving surface roughness of less than 25~nm. All electrodes were cleaned with chemical solvents, according to a previously established protocol~\cite{Malabaila2014Cleaning}.
The samples were not treated with high temperature and shallow grooves can still be observed after diamond machining on the surface, as well as individual copper grains with sizes of 20-80~$\mu$m. 
All pairs of electrodes have the same geometry, with the high field area defined by the 40~mm anode diameter.
The electrodes were separated by a ceramic spacer (Al$_2$O$_3$). The dimensions of the spacer and electrode, which are critical for maintaining the size of the interelectrode gap, were machined to a tolerance of 1 $\mu$m to ensure a high degree of parallelism, thus achieving a uniform vacuum gap between the electrodes. See Table~\ref{tab:electrodes} for details.

\begin{table*}[htbp]
\caption{\label{tab:electrodes}Common atributes for electrodes used in the study. The specified design dimensions refer to Fig.~\ref{fig:electrodes_schema}.}
\centering
\begin{tabular}{c | c | c | c | c | c| c}\hline
Electrode & Spacer   & Spacer & Diameter & Diameter & Gap size  & Gap size   \\
material  & material & height & full sample & high-field region & reference region & high-field region  \\ \hline\hline
 OFE Copper & Al$_2$O$_3$       & 20.06   & 60 & 40 & 2.06 & 0.06 \\
 (UNS~C10100) & (DEGUSSIT AL23) &   [mm]  & [mm] &  [mm] & [mm] & [mm]  \\
\hline
\end{tabular}
\end{table*}

\subsection{Conditioning}

The electrode set Cu007@300K was conditioned at CERN's DC test system~\cite{Profatilova2020BDSystem,Profatilova2019Behaviour}, while the sets Cu038@30K and Cu052@10K at Uppsala's cryogenic test system ~\cite{Jacewicz2020}. The systems share the design of the high voltage (HV) scheme and the conditioning procedure that was developed for the RF and DC tests for CLIC project~\cite{Profatilova2019Behaviour, Profatilova2020BDSystem, Saressalo2020Classification}. The cryogenic system in Uppsala, however, can cool the samples down to 4~K for the tests. 

During the conditioning, HV pulses with 1~$\mu$s width and 1~kHz repetition rate were used for the conditioning at cold and 2~kHz for Cu007@300K. Lower repetition rate at cold was used to maintain the stable temperature. The change in the repetition rate is not expected to have any significant impact on the experiment~\cite{saressaloEffectDcVoltage2020}. 
Condition tests were performed using breakdown rate feedback mode, which means that the HV was ramped up or down over time while monitoring the breakdown rate to keep it at 1$\times10^{-5}$ BD / pulse; see~\cite{Profatilova2019Behaviour} for details. The breakdown rate is calculated using an exponentially weighted moving average method over a 1 million pulses window.
Each electrode set was conditioned only at a single temperature, the Cu007@300K at room temperature, while the Cu038@30K and Cu052@10K were directly cooled down to 30~K and 10~K respectively for the test. 

\subsection{Analysis}

To better understand the effect of the conditioning with high electric field, different analysis methods were used at distinct temperatures to search and analyze the modification in the sub-surface regions and on the surface of the electrodes. The methods include high-resolution light microscopy as well as Scanning Transmission Electron Microscopy (STEM). Even though the breakdown reaction is presumably initiated from the cathode, a plasma channel is formed between the electrodes during the discharge phase, and both anode and cathode surfaces are affected. Therefore, analyses of both electrodes are of interest.

\subsubsection{Large scale structures and breakdown spot matching}
\label{sec:matlab}
\begin{figure}[htbp]
\centering
\includegraphics[width=0.99\linewidth]{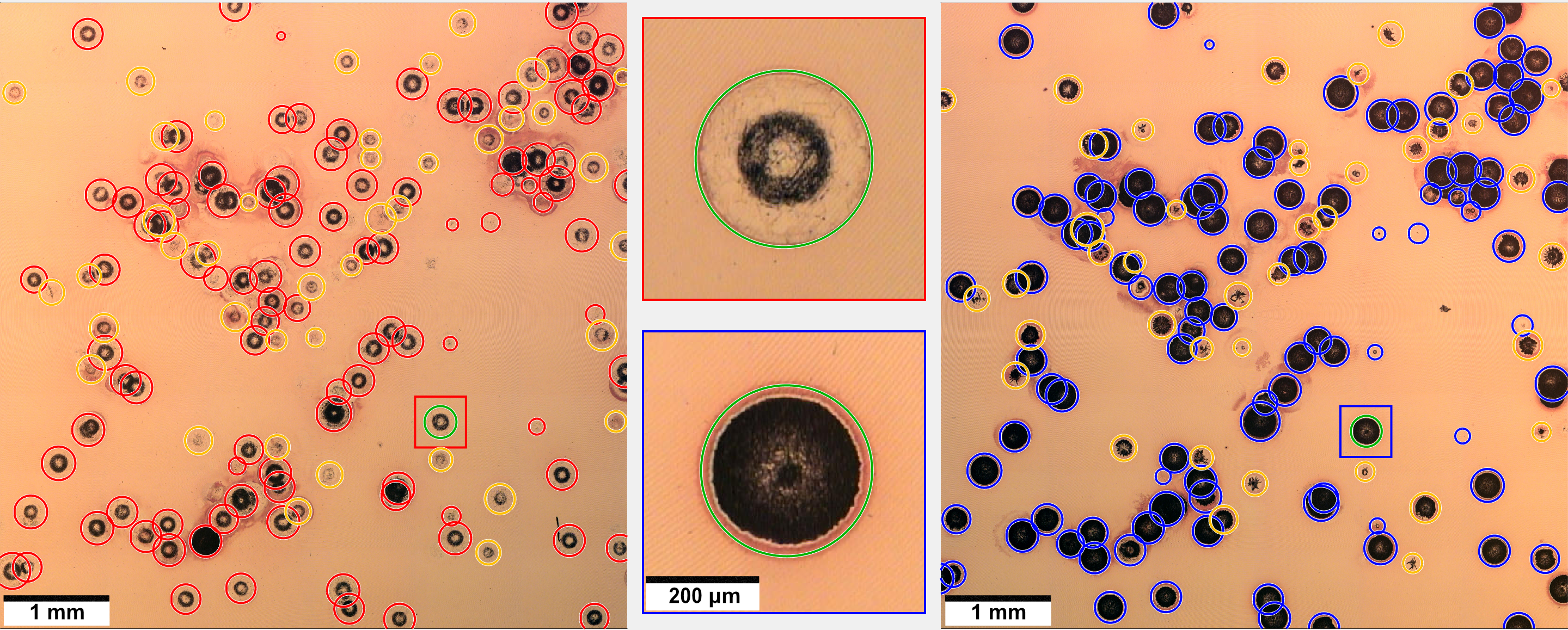}
\caption{The figure presents semi-automatic breakdown counting software with anode-cathode breakdown-site matching. The anode is on the left, the cathode is on the right. In the middle, a zoom of the selected site on the anode (top) and cathode (bottom) is shown. Presented data come from Cu038@30K set.}
\label{fig:MATLAB}
\end{figure}

Global microscopic observations of the breakdown features of the anode and cathode were done using SmartScope FLASH~200 light microscope at Uppsala University. First, the corresponding areas on anode and cathode were scanned with$~$1~$\mu$m/px resolution. The single scans were then stitched together to form a larger map. The larger maps were then analyzed using a dedicated MATLAB script, similar to the one described in~\cite{Profatilova2020BDSystem}. An automatic circle-finding routine was used at the initial stage to create an array with the coordinates of the recognized surface features, with a first estimate for the circles' radii. 
 A link between the anode and cathode images and coordinates was done semi-manually and resulted in a set of the transformation factors (a scale, a rotation angle and translation x and y positions) that allowed for direct comparison of the features on anode and cathode for the same breakdown, see Fig.~\ref{fig:MATLAB}. 

\subsubsection{Sub-surface analysis}
\label{sec:method:STEM}

\begin{figure}[h]
\centering
\includegraphics[width=0.90\linewidth]{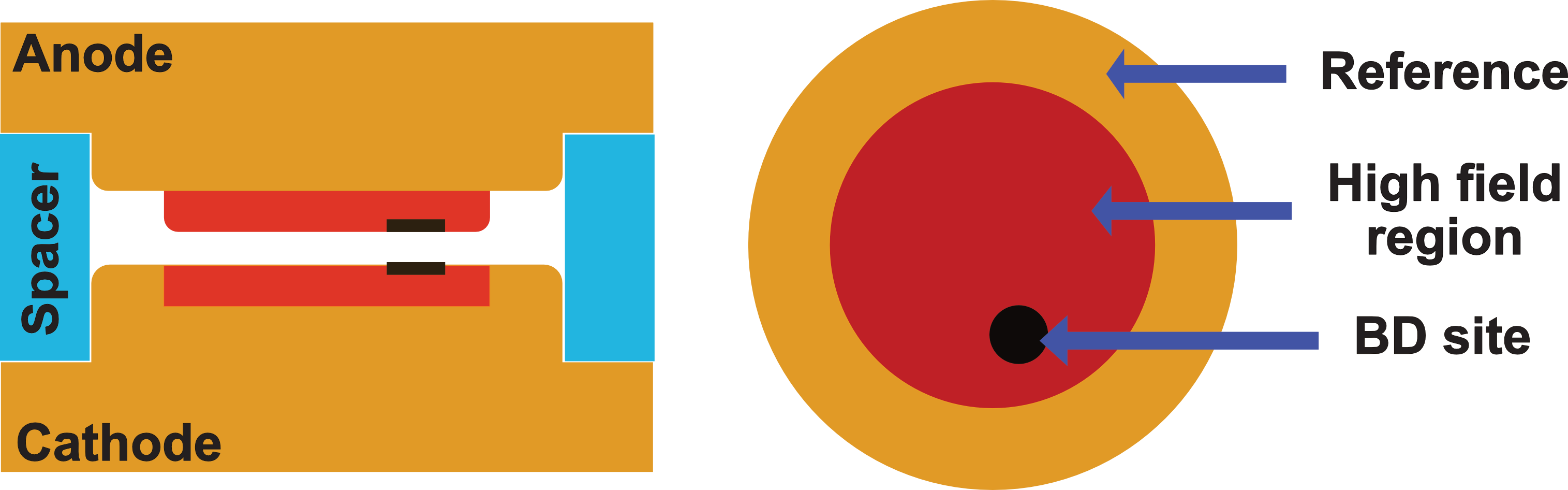}
\caption{\label{fig:electrodes_schema} A schematic view of the electrodes under test with definitions of different regions mentioned in the text (valid for anode and cathode): {\it reference} - the region with negligible electrical field (orange) with the interelectrode gap $>$2~mm, {\it high field} region (red) with the interelectrode gap 60~$\mu$m and breakdown site (small black circle)
Ceramic spacer ring is indicated in blue. The picture is not up to scale. On the left: view from the side, on the right: top view.
}
\end{figure}

The conditioned hard copper samples were analyzed with a high angular annular dark field (HAADF) STEM imaging to identify variations induced in their sub-surface structure at field exposure. This was done with Scanning-Transmission Electron Microscope (S)TEM Tecnai~F20~G2 operated at 200~kV and equipped with Fischione HAADF STEM detector. 
For comparison, cross-sectional lamellas were extracted with a Focused Ion Beam (Dual Beam FIB Helios Nanolab 460F1) from the reference regions and the regions that were exposed to the highest electrical field. The reference region is the region that underwent the same initial treatment, but was not exposed to high fields (see Fig.~\ref{fig:electrodes_schema} and Table~\ref{tab:electrodes} for the details). 
The analysis of the STEM data included counting dislocations and calculating a local dislocation density.

\section{Results}
\label{sec:results}

\subsection{High-voltage BD behavior during conditioning}

\begin{figure}[h]
\centering
\includegraphics[width=0.90\linewidth]{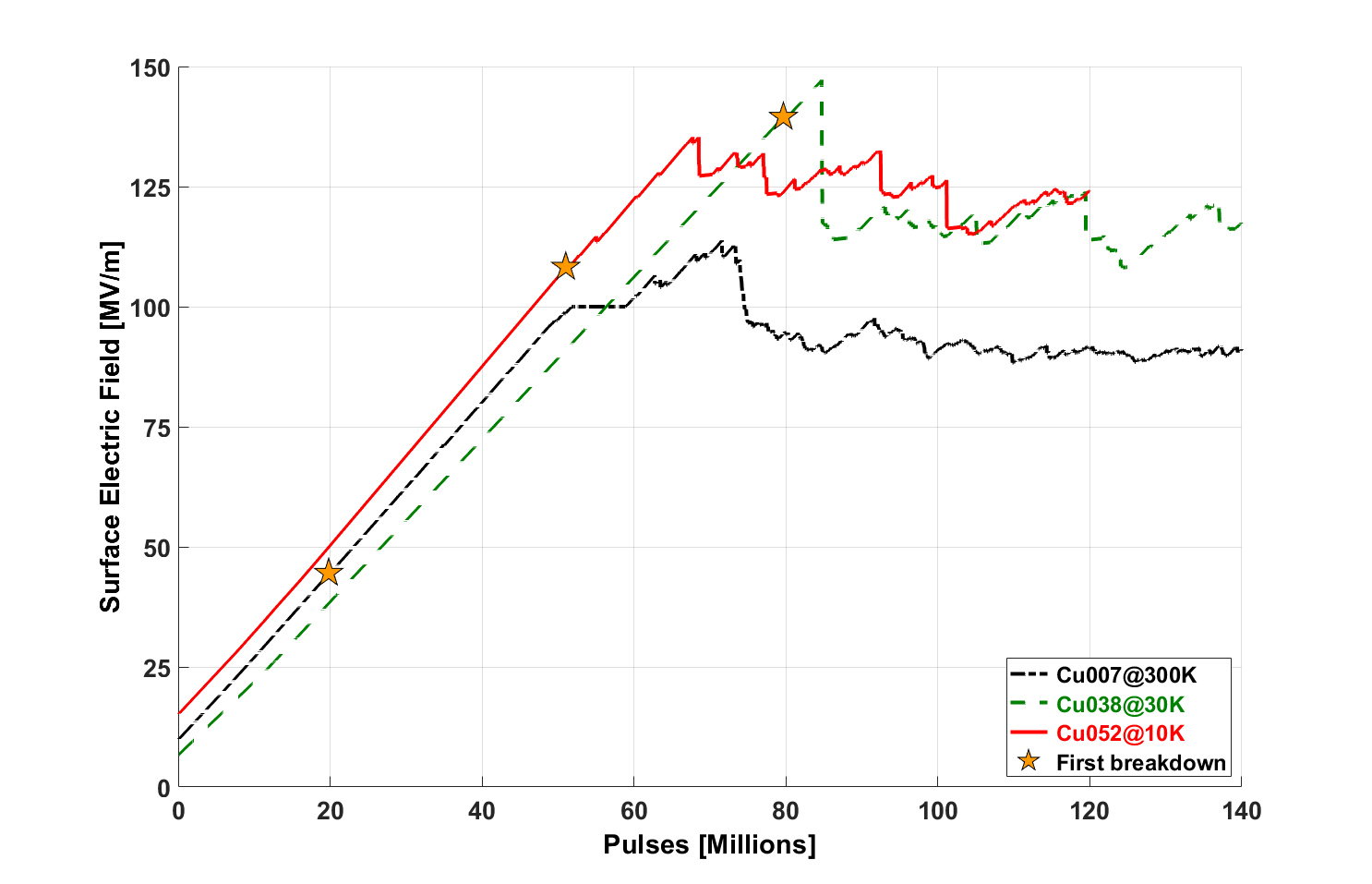}
\caption{\label{fig:conds_comparison} Comparison of conditioning curves for sample: Cu007@300K (black dashed-dotted line), Cu038@30K   (green dashed line) and Cu052@10K (red solid line). A yellow star on the conditioning curve indicates the pulse number corresponding to the occurrence of the first breakdown.}
\end{figure}

Typically, during the conditioning, the samples should reach stable field level with a BD rate below 1$\times10^{-5}$ BD/pulse.
This value, calculated by averaging over the last 10 million pulses, is called the saturation field.
Due to the conditioning process's stochastic nature, the fields can fluctuate at different levels before the saturation field is reached. These fluctuations are of interest, and one of the factors that can describe them is the overall maximum attained field and the number of pulses and BDs needed to reach that field. 
The comparison of the conditioning curves for all the samples is shown in Fig.~\ref{fig:conds_comparison} and the main information extracted from the measurements is given in Table~\ref{tab:conditioning}. The gap size was determined using capacitance measurements of the cooled system and was the primary source of uncertainties listed in the table.
The automatic conditioning algorithm dictates the constant slope visible during the first part of the conditioning, as voltage is ramped at a constant rate. 
 
 \begin{table*} [h]
 \caption{\label{tab:conditioning} Main parameters for the tests.}
     \centering
     \begin{tabular}{c|c|c|c|c|c|c} \hline 
   Electrodes&  Temperature &  Gap &  First BD E$_{field}$ &  Max E$_{field}$ &  Saturation E$_{field}$ & Nb of BDs to max E$_{field}$\\
      &  [K]& [$\mu$m]&  [MV/m]& [MV/m]&  [MV/m]& \\ \hline\hline
          Cu007@300K&  300&  60$\pm$3&  46$ \pm$3&  114$\pm$6&  90$\pm$5& 94\\
          Cu038@30K&  30&  59$\pm$3&  140$\pm$8&  147$\pm$8&  120$\pm$7& 7\\  
          Cu052@10K&  10&  59$\pm$3&  108$\pm$5&  135$\pm$7&  123$\pm$7& 27\\\hline
     \end{tabular}
     
 \end{table*}

\subsection{Characteristics of common surface features of the breakdown sites}
BD events lead to the formation of crater-like features on metallic surfaces on both sides of the event~\cite{anders_cathodic_arcs}. 
The cathode and anode surfaces were analyzed using microscopy to characterize these typical features, and the damaged areas were classified. Specific events were identified using the procedure described in section~\ref{sec:matlab}.

\begin{figure*}[h]
    \centering
\includegraphics[width=\textwidth]{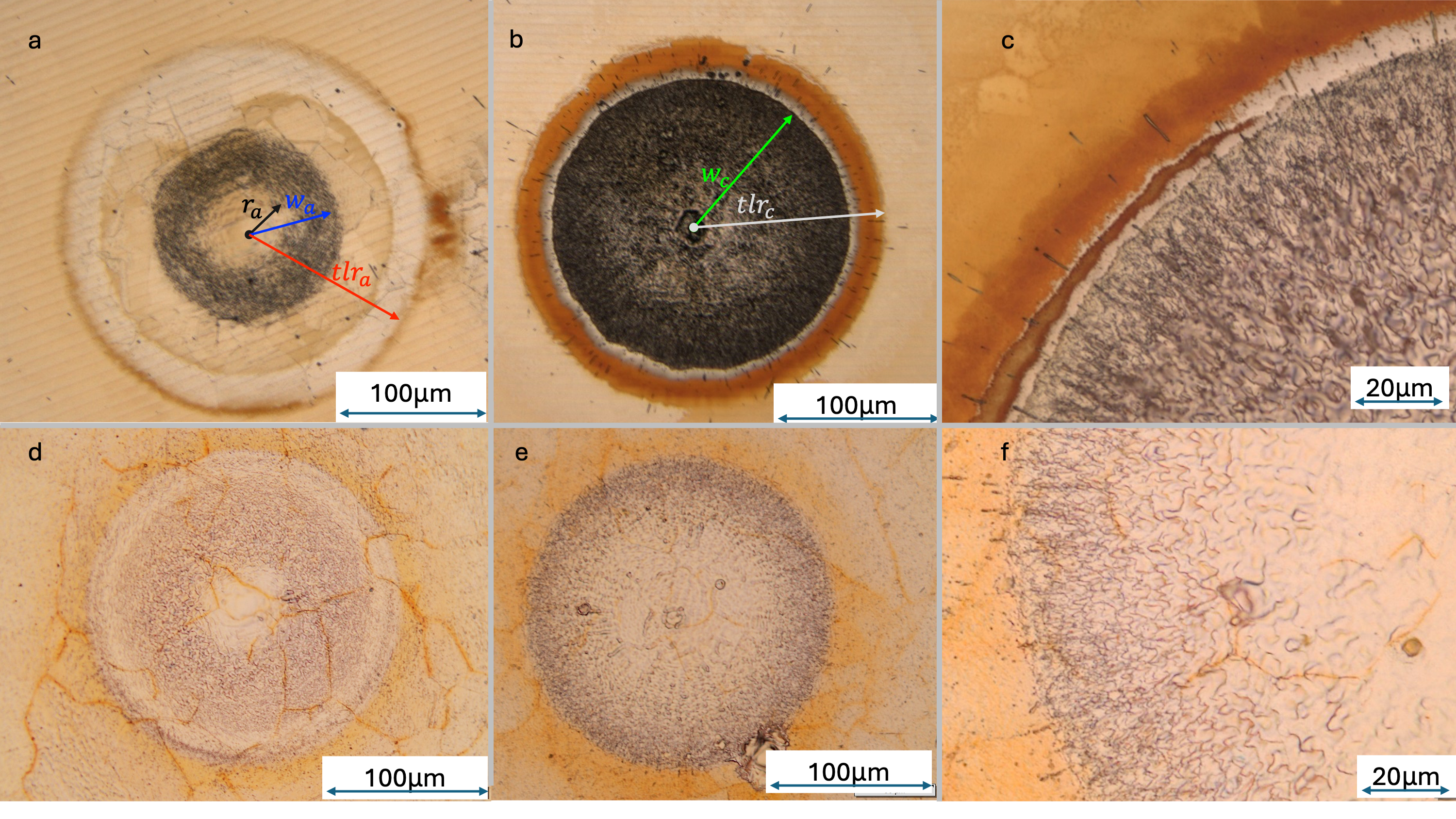}
     \caption{Typical BD features on sample Cu038@30K - at the top row : (a) Anode side: a round-shaped anode spot composed of a reflective central part (with radius of $r_a$) surrounded by a rough circular region (with a radius of $w_a$). The whole spot is surrounded in a lighter circular region (radius of $tlr_a$). (b) Cathode side: a cathodic spot composed of a large, dark, rough circular region (radius of $w_c$) and a thin lighter outer ring (radius of $tlr_c$). (c) Zoom in on small jets at the edge of the dark region (around $w_c$). These were observed to cross even to the outer lighter region. Bottom row has the same pictures but for sample Cu007@300K with pictures of a spot on the anode, cathode, and zoomed in on the margins of the cathodic spot, as sub-figs d,e,f correspondingly.}
        \label{fig:Typical_BD}
\end{figure*}

Fig.~\ref{fig:Typical_BD}(a) shows a typical "crater" left after a breakdown on an anode, with a halo around the center most likely due to a capillary wave formed in a molten state. It is maximum 0.5~$\mu$m deep, with shallow machining grooves still visible throughout the spot~\cite{Saressalo2021PhDThesis}. A typical round-shaped anode BD feature is composed of a reflective central part, with a radius of $r_a$, surrounded by a rough circular region, with a radius of $w_a$. This outer region is surrounded by a circular region of lighter shade, with a radius of $tlr_a$. 

A typical cathode BD feature has a large uniform rough circular region, with radius  $w_c$, and a thin lighter outer ring, with radius of $tlr_c$, as illustrated in Fig.~\ref {fig:Typical_BD}(b). 
The cathode BD feature usually has a crater with  1~$\mu$m depth in the center ~\cite{Saressalo2021PhDThesis}. On the cathode side, the machining grooves are not visible at the BD site.  WE speculate that this is due to higher energy deposition on this side, leading to higher surface temperatures and deeper damage. 
We found emitted jets of copper in random radial directions surrounding the BD crater on this side, Fig.~\ref{fig:Typical_BD}(c), while similar jets are rarely observed on the anode side. 
These features were reproduced independently of the conditioning temperature, as demonstrated in the lower row of Fig.~\ref{fig:Typical_BD}.

\cite{}

\subsection{Star-like BD features}
\begin{figure*}[htbp]
\centering
\includegraphics[width=0.8\linewidth]{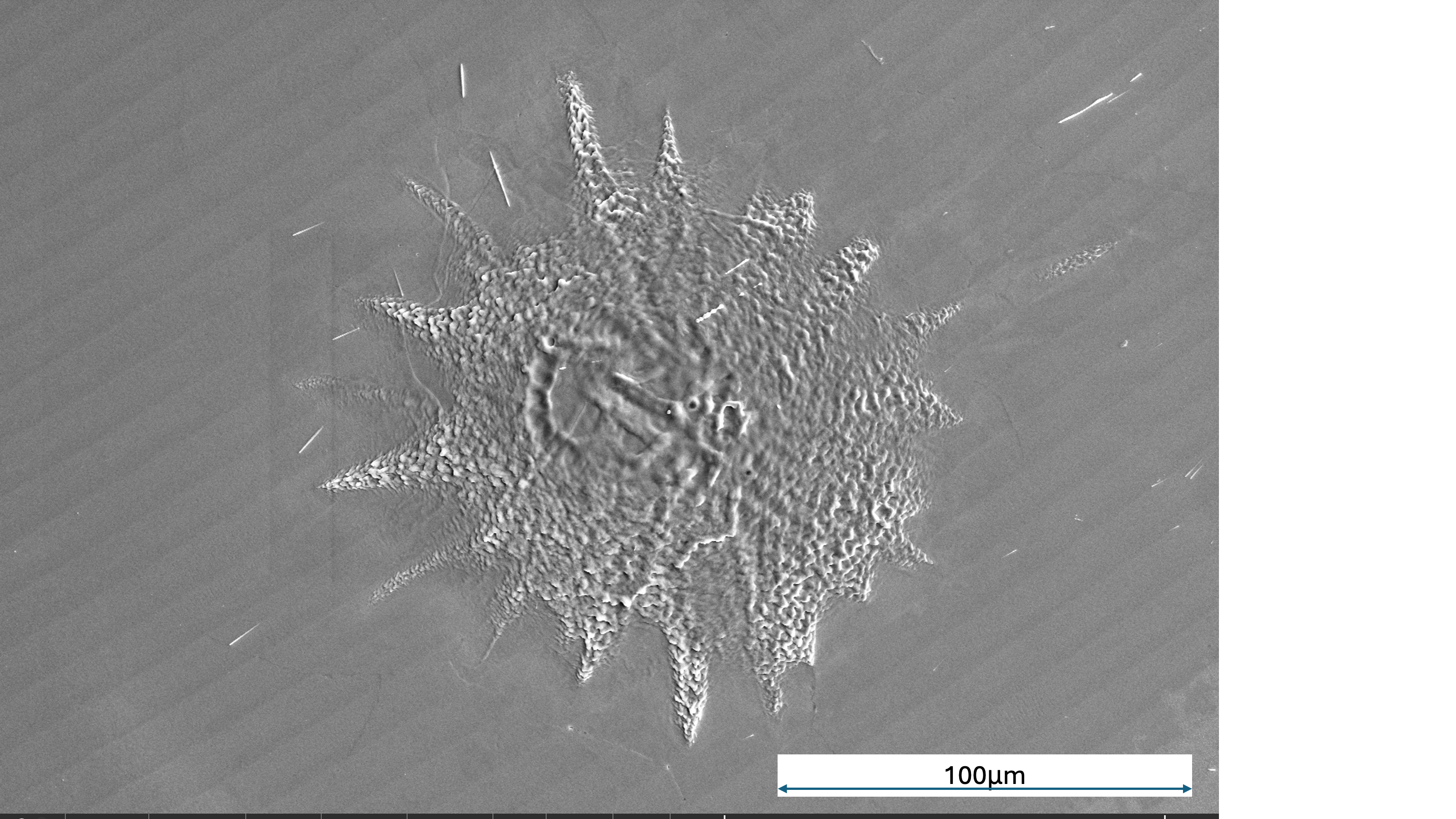}
\caption{\label{fig:inna02star} SEM image of a star-like BD site on the cathode Cu038@30K. The details of the star-like deposited material are clearly visible; The machining grooves, resulting from diamond-turning, are visible only out of the material deposition zone. 
The lighter halo region observed in the optical microscopy, radius of $tlr_c$ in Fig.\ref{fig:Starlike_BDs}(d), is not clearly visible here.  
}
\end{figure*}

A number of atypical features of BD were discovered in the samples tested at cold temperatures. 
The spots are characterized by a very shallow depth with a star-like shape instead of a circular one.
The star-like BD features occurred only on the cathode. See the SEM image on Fig.~\ref{fig:inna02star} for the star-like BD morphology details. 

\begin{figure*}[htbp]
\centering
\includegraphics[width=1\linewidth]{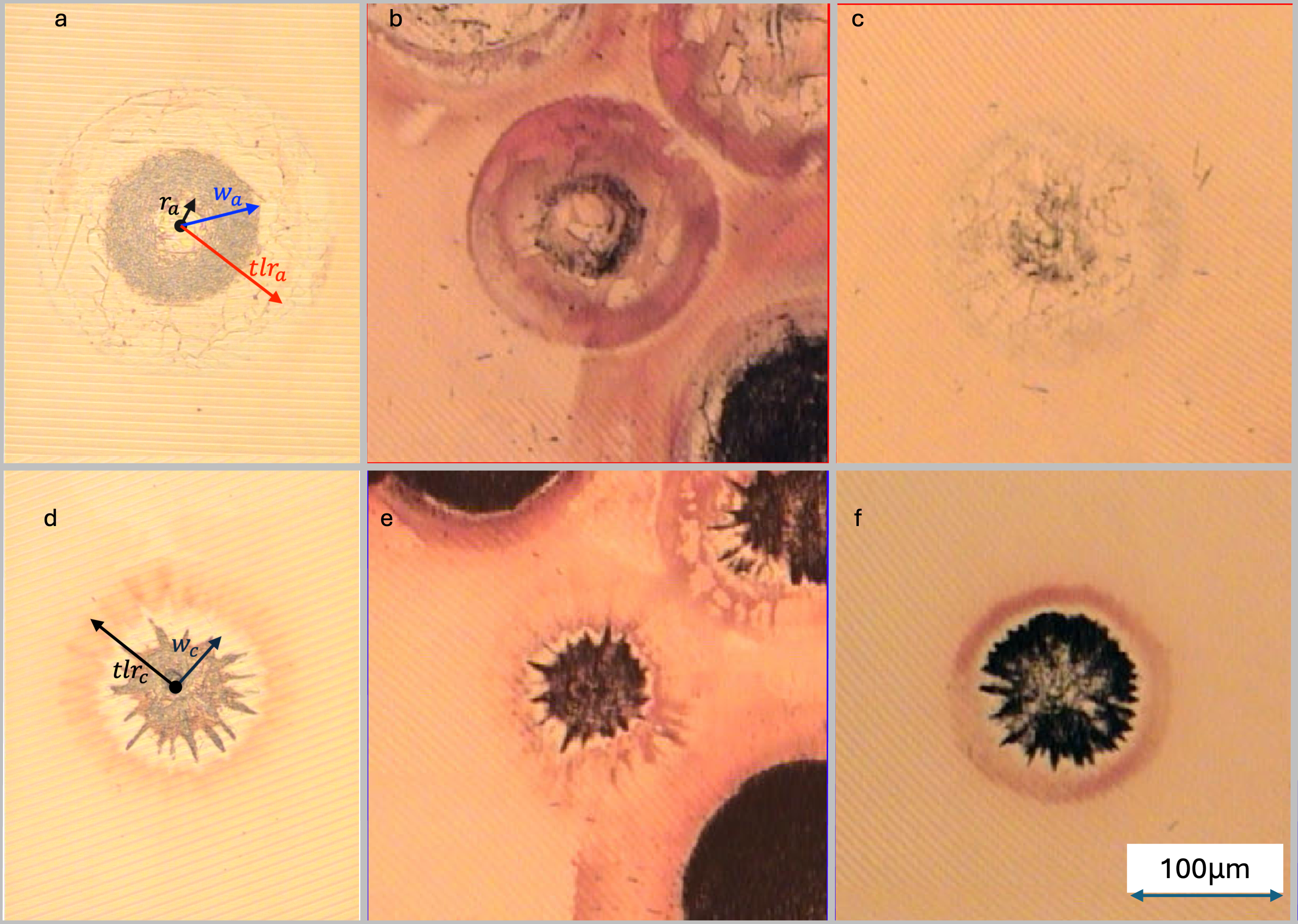}
         \caption{Breakdown features on the anode, top row, (a-c), and corresponding atypical feature found on the cathode (d-f). On the left, a and c, the definitions of the radii is similar to the definitions in figure \ref{fig:Typical_BD}, with the main difference that the rough circular region ($w_c$) now marks the outer circumference of the star-like rough region. 
       }
        \label{fig:Starlike_BDs}
    \end{figure*}

More examples of optical microscopy pictures of atypical BD features are shown in Fig.~\ref{fig:Starlike_BDs} for sample Cu052@10K. 
For each star-like spot on the cathode side, we matched the corresponding spot on the anode side (on the same in-plane location). 
These are shown as couples with top and bottom rows showing features from the same breakdown spot-pair on the anode and cathode, respectively.
In a similar manner to the characterization of the typical spots, we also defined here the average radii of the different features.
 Figs.~\ref{fig:Starlike_BDs}(a) and~\ref{fig:Starlike_BDs}(d) demonstrate how the regions for the BD features are defined. Note that since no start-like features appear on the anode side, the definition of radii in Fig.~\ref{fig:Starlike_BDs}(a) is similar to that in the typical spot (see Figs.~\ref{fig:Typical_BD}).
The BD features on the anode side look similar and retain the visibility of the machining grooves in all regions except the rougher disk between $r_a$ and $w_a$.
The spots also have a similar size distribution, although the details for these were less pronounced.
This suggests that they were shallower, but such a hypothesis needs to be further pursued using dedicated cuts of specific spots.
 
\subsection{Statistical analysis}
The high-resolution scans of the cathode and anode surface were analyzed with the software mentioned in section~\ref{sec:matlab}, matching the corresponding BD spot with close to 100\% accuracy. Afterward, the regular and atypical features were characterized by looking for similarities and differences, and quantifying the relationship between the temperature and the number of different classes of BDs.

\subsubsection{Regular features}

The ratios between different anode and cathode features were studied in detail, mostly for the electrode set Cu038@30K   and Cu052@10Kand summarized in Table~\ref{tab:ratios}. For Cu007@300K, only average radius sizes are presented, as the craters on the cathode and anode were not linked to each other during analysis, and star-like features were not found in the room-temperature test. 

The ratios between the cathode and anode radii found for the cold sets are very similar, agreeing within the uncertainties. For example, the ratio for the rough dark circular regions, $w_c/w_a$, was found to be $1.6\pm0.3$ for Cu038@30K   and $1.5\pm0.2$ for Cu052@10K (also see Fig.~\ref{fig:hist_radiuses}(a)) and the ratio between the radii of thin light rings of cathode and anode, $tlr_c/tlr_a$, was found to be around $1$ (shown in Fig.~\ref{fig:hist_radiuses}(b)).

\begin{table*}[htbp]
\caption{\label{tab:ratios}Average radii and ratios between sizes of BD features.}
\centering
\begin{tabular}{ c | l | c | c | c } \hline
 Type & Value & Cu007@300K& Cu038@30K  & Cu052@10K\\ \hline \hline

 \multirow{6}{*}{Regular} & $r_a$ & 20 $\pm$ 4 & 21 $\pm$ 4  & 23 $\pm$ 3 \\
 & $w_c/w_a$ & & 1.6 $\pm$ 0.3  & 1.5 $\pm$ 0.2 \\
  & $tlr_c/tlr_a$ && 1.0 $\pm$ 0.1 & 1.0 $\pm$ 0.1\\
  & $tlr_a/w_a$ & & 1.9 $\pm$ 0.3 & 2.0 $\pm$ 0.2 \\
  & $tlr_c/w_c$ && 1.2 $\pm$ 0.2 & 1.3 $\pm$ 0.2 \\
  & Average radius on cathode ($w_c$), $\mu$m & 101 $\pm$ 20 & 125  $\pm$ 16&  111 $\pm$ 16 \\
  & Average radius on anode ($w_a$), $\mu$m& 73 $\pm$ 12 & 78 $\pm$ 11 & 76 $\pm$ 10 \\\hline
 \multirow{6}{*}{Star-like} & $w_c/w_a$ && 1.6 $\pm$ 0.5 & 1.5 $\pm$ 0.1 \\
  & $tlr_c/tlr_a$& & 1.0 $\pm$ 0.1 & 1.0 $\pm$ 0.1\\
  & $tlr_a/w_a$  && 1.7 $\pm$ 0.4 & 1.9 $\pm$ 0.2\\
  & $tlr_c/w_c$ && 1.1 $\pm$ 0.2 & 1.3 $\pm$ 0.1 \\
  & Average radius on cathode ($w_c$), $\mu$m & -- & 109 $\pm$ 25 &  111 $\pm$ 14 \\
  & Average radius on anode ($w_a$), $\mu$m & -- & 70 $\pm$ 28 & 77 $\pm$ 8 \\\hline
\end{tabular}
\end{table*}

\subsubsection{Atypical features}

The star-like features were observed exclusively on the cathodes of samples tested at cryogenic temperatures. These breakdowns were counted and correlated between both electrodes, similar to the procedure used for regular sites. The Table~\ref{tab:statistics} presents statistics comparing the number of atypical features detected at 30~K and 10~K.

\begin{table*}[htbp]
\caption{\label{tab:statistics} Main statistical information about the microscopic images.}
\centering
\begin{tabular}{ c | c | c | c }\hline
 Samples & Temperature & Nb of BDs & Nb of star-like pairs (and percentage) \\ \hline
Cu038@30K & 30 K & 145 & 37 (26 $\pm$ 5 $\%$) \\
Cu052@10K& 10 K & 280 & 149 (53 $\pm$ 7 $\%$) \\\hline
\end{tabular}
\end{table*}

\begin{figure*}[htbp]
\centering
\includegraphics[width=1\linewidth]{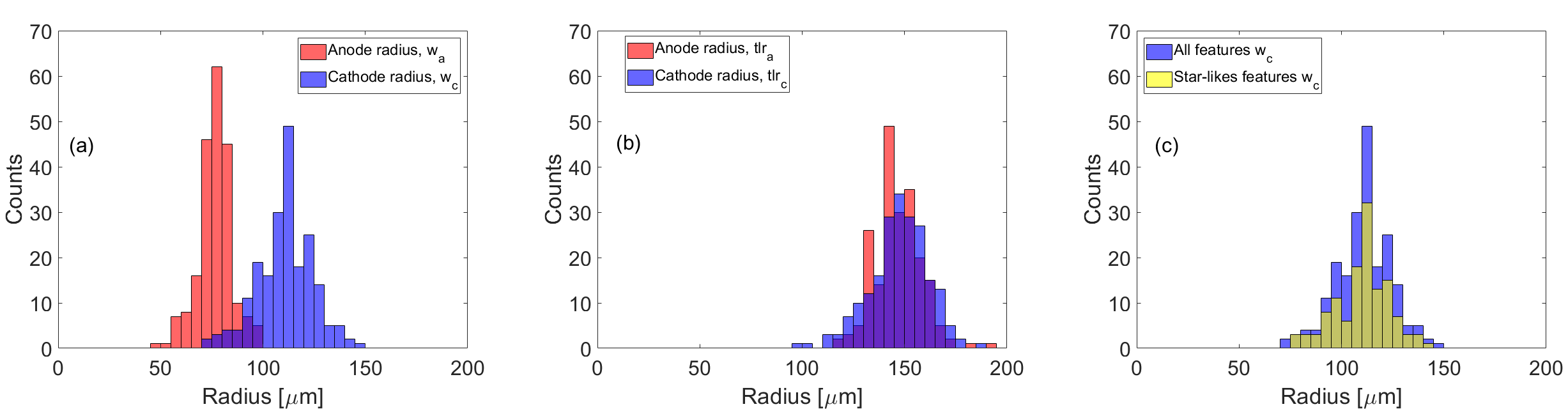}
\caption{Comparison of radii for Cu052@10K: (a) anode vs cathode for all BD features (dark rough circular region), (b) thin light rings for all BD features for anode and cathode, (c) star-like and the all BD feature on the cathode (dark rough area).}
        \label{fig:hist_radiuses}
\end{figure*}

Following the initial analysis, a circle was fitted to each star-like spot to estimate its size. A comparison of the radii of anode and cathode spots is presented in Table~\ref{tab:ratios}. 
Fig.~\ref{fig:hist_radiuses} includes results from all the features found on the Cu052@10K surface where anode-cathode spots could be matched, including star-like spots. 
The results show that the distribution of the BD characteristic radii on the anode and cathode varies depending on the region considered. 
The difference is most noticeable for the rough circular regions, with 70 - 150~$\mu m$ for cathode versus  50 - 100~$\mu m$ for anode. Meanwhile, for the same BD features, the radii for the regions of thin lighter outer rings ($tlr_a$ and $tlr_c$) are nearly equal with a distribution between 100 - 200~$\mu m$ (Fig.~\ref{fig:hist_radiuses}(b)). 
The radius distribution of the rougher regions for circular versus star-like spots is similar, as shown in Fig.~\ref{fig:hist_radiuses}(c).  

\subsection{Sub-surface features} 
\label{sec:results:STEM}
The conditioned copper samples were analyzed after field exposure to identify variations in their sub-surface structure. This was done by creating an HAADF STEM picture of cross-sectional lamellas using field ion beam lithography.
Lamellas taken from field-exposed regions were compared to lamellas from the reference regions, see Fig.~\ref{fig:electrodes_schema}.
 
\begin{figure}[htbp]
\centering
\includegraphics[width=0.99\linewidth]{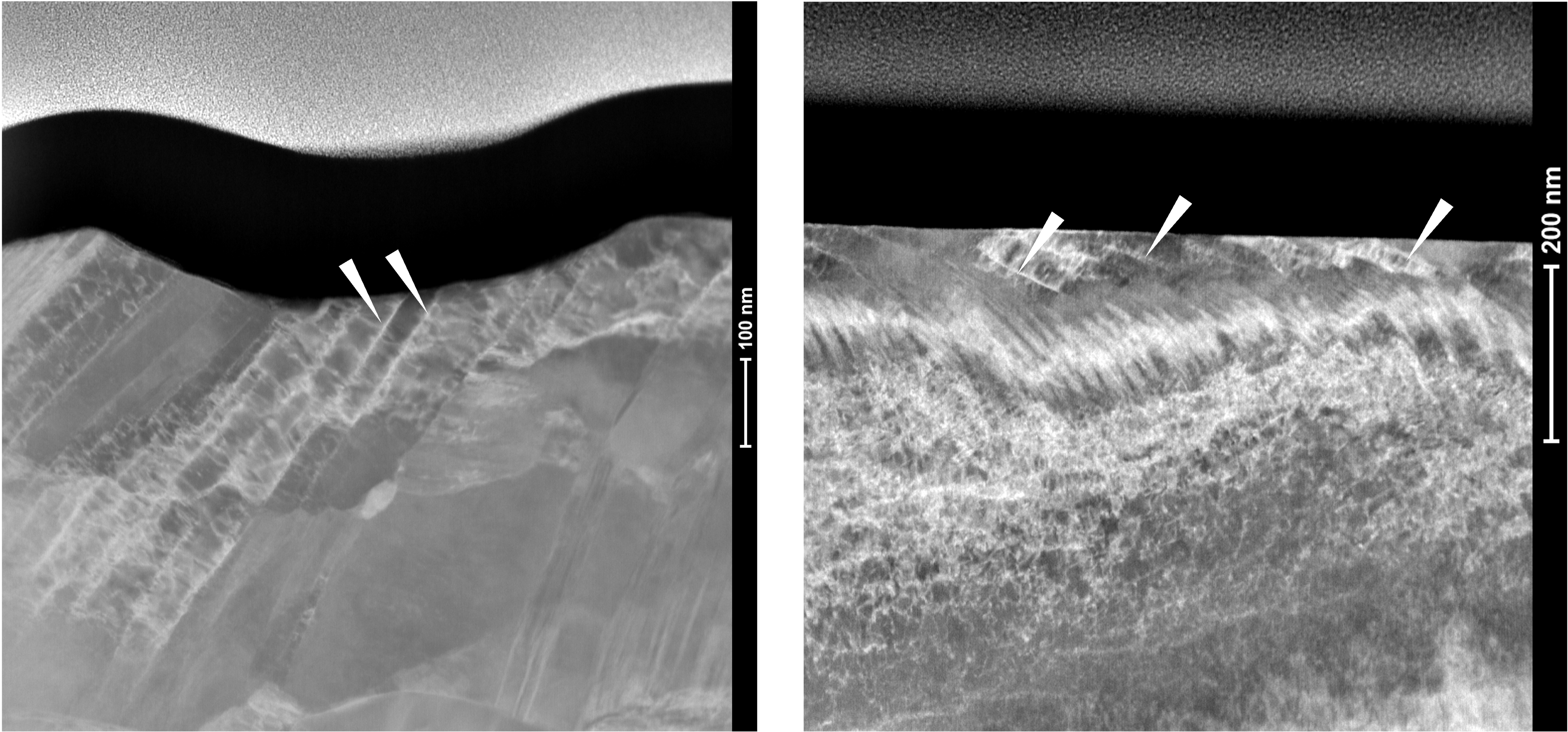}
\caption{Cathode structure below the surface in reference regions on the side of the sample, which were not exposed to high electric fields. Images are from cathodes conditioned at room temperature (left, sample Cu007@300K) and 30~K (right, sample Cu038@30K). White arrows mark dislocation walls of sub-micron domains introduced by the diamond turning of cathode surfaces. Sample Cu007@300K was cleaned by chemical etching, leading to a non-uniform removal of the top 100~nm.} 
\label{fig:huji1}
\end{figure}

\begin{figure}[htbp]
\centering
\includegraphics[width=0.99\linewidth]{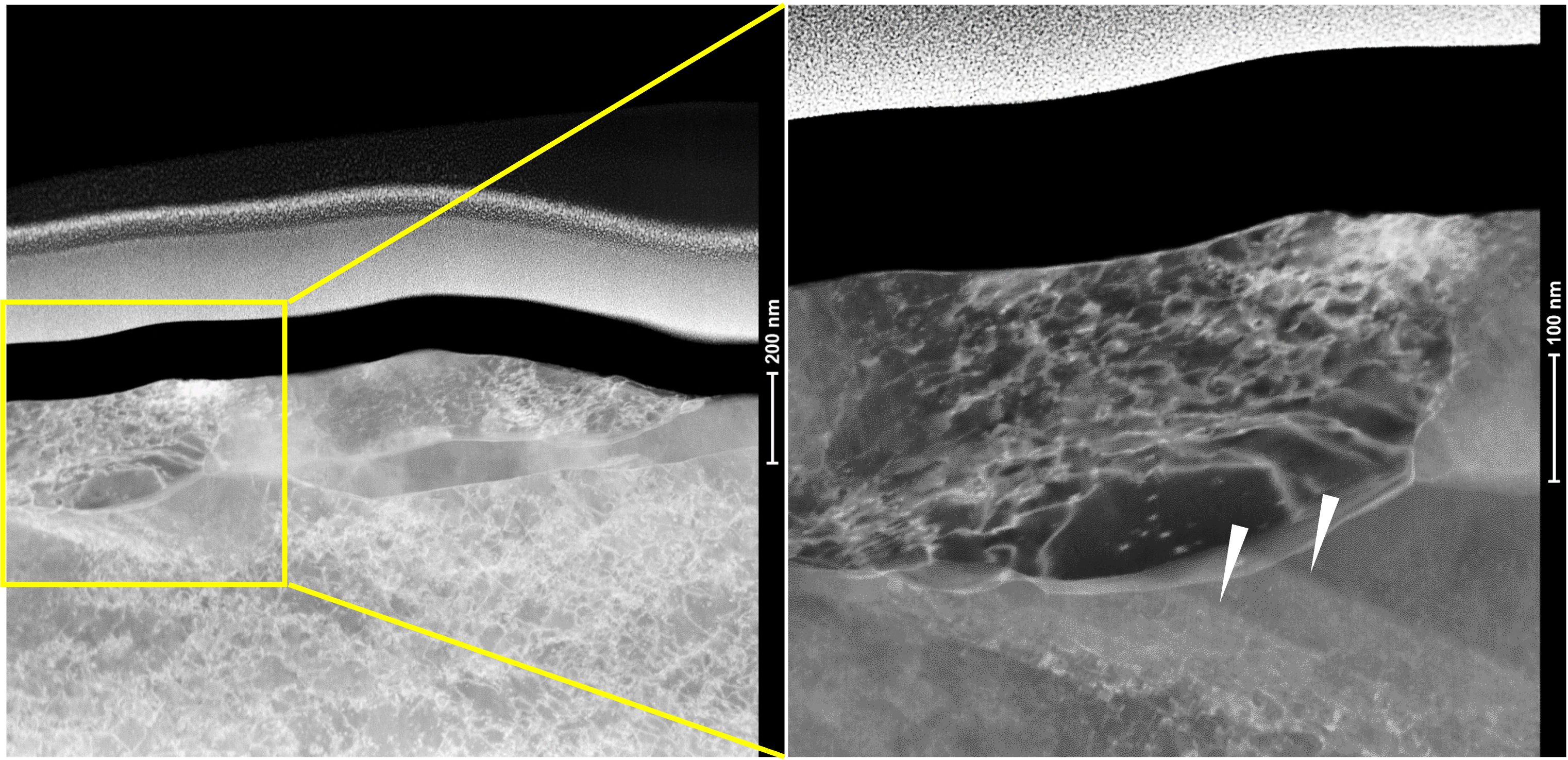}
\caption{Sub-surface structure in the field-exposed region of the sample Cu007@300K. The right panel is a magnified view of the region marked on the left one. White arrows mark the remaining dislocation walls.}
\label{fig:huji2}
\end{figure}

\begin{figure}[htbp]
\centering
\includegraphics[width=0.99\linewidth]{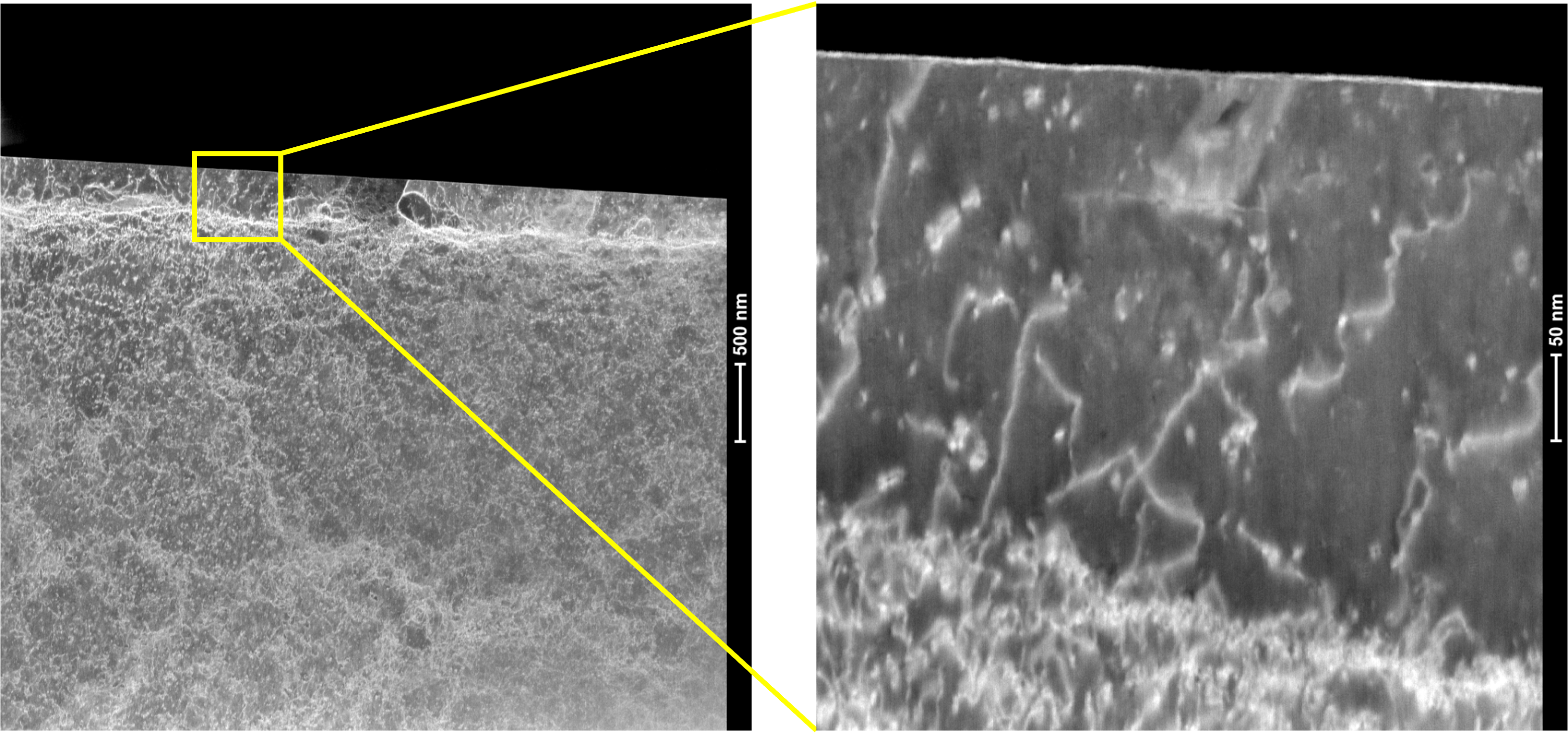}
\caption{Sub-surface structure in the field-exposed region of the sample Cu038@30K. The right panel is a magnified view of the region marked on the left one, demonstrating the formation of a top layer with a significant reduction in the dislocation density.} 
\label{fig:huji3}
\end{figure}

The reference sub-surface regions of cathodes Cu007@300K and Cu038@30K   contained a rather high density of structural defects induced by the diamond turning (Fig.~\ref{fig:huji1}). 
Although grain structures seem to differ between samples, this can be attributed to the low-temperature treatment sample Cu038@30K   was exposed to, which generated a significant thermal load and is consistent with previous observations of grain refinement and strong surface effects on dislocation structure~\cite{Sonar_2018,Shengquan_2019}. In both samples, grain structure and dislocation walls are observed to be continuous up to the surface. 
However, clear variation in structure can be observed in the conditioned regions. 
Conditioning in both cases significantly reduces the number of dislocation walls (Fig.~\ref{fig:huji2}). 
This effect is more pronounced in the cold-conditioned sample Cu038@30K, where the formation of a dislocation denuded zone close to the surface is clearly visible (Fig.~\ref{fig:huji3}). 

\subsection{Hardness}

In addition to microscopy, Vickers hardness was measured for the copper sample conditioned at 30~K.
Hardness was measured at two types of locations: area exposed to high field during conditioning but without visible breakdown site, and inside the breakdown spot, within the perimeter of molten copper. 
Measurements were done with a microhardness tester Duramin-40M3 (Struers) under 20~\textit{gf} load and 10~seconds dwelling, and were repeated at 10 separate locations for each location type. 
Measurements were done separately at the center of the sample, up to a distance of 1~cm from the center of the electrode, within 1~cm from the outer region of the exposed area, and near the electrode edge, which was not exposed to high field, see Fig.~\ref{fig:electrodes_schema}. The results are presented in Table~\ref{tab:hardness}. The hardness of the conditioned regions varied between 90 and 110 (HV) with lower values attained at BD sites and higher values in between these sites. 
These values are within the range of hardness attained at a reference outer region of the sample, which was not directly exposed to the same high fields. We note that the lower range of values was measured at sites where the surface was modified due to thermal effects, and cross-sectional microscopy showed the formation of a recrystallized melted layer. 
 All these values are considerably higher than $47 \pm$ 3~(HV)~\cite{Popov2021}, measured on a heat-treated sample of the same material, resulting in 'soft' copper with large grain size. See, for example, reference~\cite{korsback2020} for a detailed description of this treatment and results of conditioning. We therefore conclude that hardness as measured by this method is not directly linked to the surface's conditioning or field retention ability.

\begin{table*}[htbp]
\caption{\label{tab:hardness}Hardness of copper, Vickers number (HV).}
\centering
\begin{tabular}{c | c | c | c | c }\hline
Reference & \multicolumn{2}{c|}{\textbf{Conditioned non-BD site}} & \multicolumn{2}{c}{\textbf{Conditioned BD site}} \\ \hline
  & center & edge & center & edge \\ \hline
 104$\pm$ 3 & 107 $\pm$ 4 & 108 $\pm$ 4 & 97 $\pm$ 6 & 93 $\pm$ 4 \\
\hline
\end{tabular}
\end{table*}

\section{Discussion}

A series of measurements were performed with three pairs of planar copper electrodes, all of which were manufactured and treated similarly. The sets underwent high-field conditioning with the identical effective gap size (60~$\mu$m) and following the same procedure until the saturation field for the set was found. The conditioning involved repeated exposure to short high-voltage DC pulses, allowing for occasional vacuum breakdown formation, resembling the operation of a high gradient RF cavity. The conditioning was performed at a single temperature, unique for each set, i.e. at 300, 30, and 10~K, respectively. 

\begin{figure*}[htbp]
\centering
\begin{subfigure}[b]{0.43\textwidth}
        \centering
        \includegraphics[width=\linewidth]{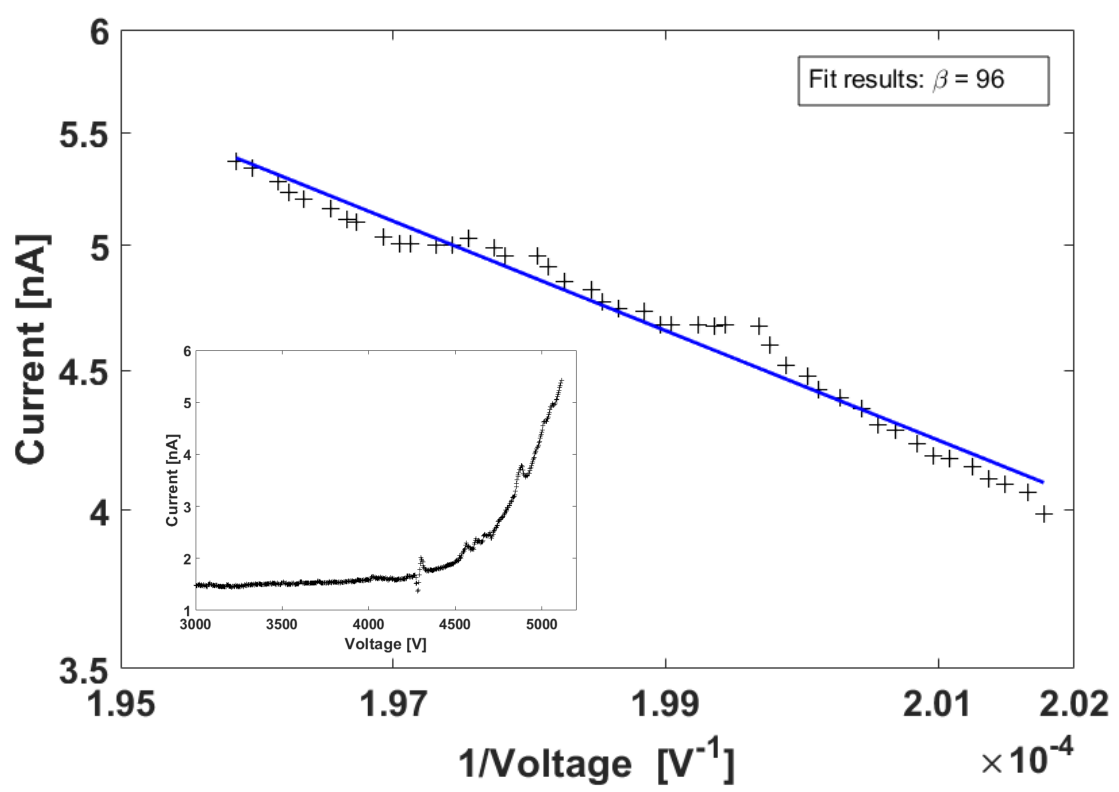}
        \caption{}\label{fig:038_beta}
\end{subfigure}
\begin{subfigure}[b]{0.53\textwidth}   
            \centering 
            \includegraphics[width=\textwidth]{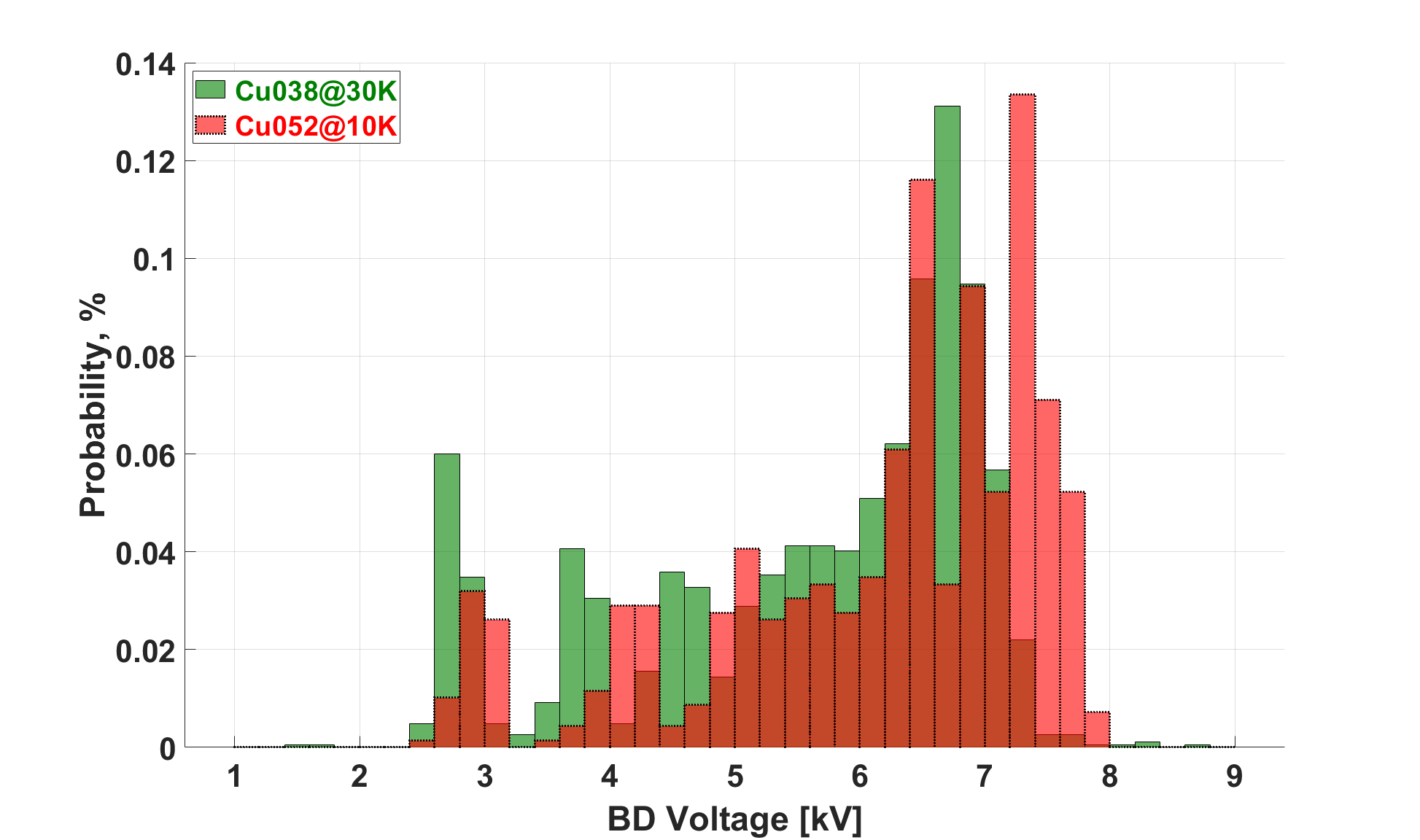}
            \caption{}\label{fig:BD_voltages}   
\end{subfigure}
\caption{(a) Fowler-Nordheim plot after field emission from Cu038@30K fitted to extract the local field enhancement factor. The inset shows the original I-V plot. (b) Histogram of BD voltages for the samples Cu038@30K and Cu052@10K tested at 30 and 10~K showing a very similar range of BD voltages.
}
\label{fig:summary01}
\end{figure*}

The conditioning at cryogenic temperatures led to a significant increase in the field holding capacity of the electrodes, in agreement with previous observations~\cite{Jacewicz2020, Cahill2018a}. 
In addition to the difference in saturation values, the field in which the initial BD event was observed increased significantly with a decrease in the working temperature of each electrode set. In the case of electrodes tested at room temperature, initial BD was observed after 20 million pulses at 45~MV/m, while the electrodes tested at 10 and 30~K had the first BD occurring after 50 and 80 million pulses at 108 and 139~MV/m, respectively (see Fig.~\ref{fig:conds_comparison}). 
This observation is consistent with the previous hypothesis that high field pulsing rather than exposure to BDs serves to condition the sample~\cite{Degiovanni2016}. 
To separate the effect of the conditioning from the hardening due to cryo-cooling we performed field emission scans after the cooling, but before the conditioning on Cu038@30K. 
To avoid BDs, the emitted current was limited to tens of nano-amperes. When the pre-breakdown emission current between the electrodes is measured as a function of the applied voltage, a so-called Fowler-Nordheim plot is obtained; see Fig.~\ref{fig:038_beta}. 
By measuring the slope of the data plot, the local field enhancement factor $\beta$ can be determined~\cite{sternFurtherStudiesEmission1997}. The macroscopic breakdown voltage $V_M$ is related to the known local critical breakdown field $E_{local}$, believed to be an intrinsic parameter of the material and found to be 10.8~GV/m for copper~\cite{descoeudres_dc_2009}. Using

\begin{equation}
E_{local} = \beta \frac{V_{M}}{d} = \beta E_{macro}
\end{equation}
where $d$ is the gap size, and using the fitted value of $\beta = 96.3$, we obtain $E_{macro} = 112$~MV/m, below the value where the first BDs occurred ($E_{macro} = 140$~MV/m). This observation suggests that conditioning might be achieved even with HV pulsing without breakdowns. However, this calls for future systematic study, varying the conditioning procedure and stages.
After the onset of BDs, the field holding slightly decreased and remained at around 120 MV/m. This deconditioning effect has been observed previously, but its underlying causes remain unclear.

To examine the variation in the sub-surface structure of the conditioned regions, we compared STEM pictures from regions that were exposed to high field, at the central region of the electrodes, to regions that were not exposed to high field, at the outer regions of the electrode. This comparison allows for the evaluation of variation due to field exposure, as samples were taken from different locations on a single surface; thus, they were exposed to a single machining and preparation process.  
These demonstrated that indeed significant plastic activity occurred in regions exposed to high electric fields, leading to a significant reduction in the dislocation density and number of dislocation walls. This initial observation is consistent with previous suggestions that conditioning is related to sub-surface dislocation activity \cite{Engelberg2018PRL,engelberg_theory_2019} but needs to be repeated more systematically. 
This is also consistent with the fact that conditioning was not observed to correlate with an increase in measured surface hardness.   
 
After high-field conditioning at warm and cold temperatures, the microstructural changes in the electrode materials at BD sites, were investigated with a light microscope and SEM. The surface of the electrodes was photographed with high resolution, allowing the identification of imprints left by the BD event on the anode and cathode sides for all the BDs.
The regular BD spots on the anode and cathode were characterized according to their morphology. 
The measured sizes of each region on each electrode agree between the sets, with the slightly larger values for the cryogenically cooled sets. The deviation between the size can be attributed to different conditioning voltages sustained at cold and warm and with this the energy available for the BD to feed the plasma.

Some of the breakdown events were accompanied by the creation of an atypical BD spot. The number of such spots increased with the reduction of the conditioning temperature, from 25\% at 30~K to 53\% at 10~K. The parameters of the conditioning process used for both sets were kept the same and both reached similar saturated voltages, illustrated in Fig.~\ref{fig:BD_voltages}. 
The presented histograms do not fully overlap, probably due to the stochastic nature of the conditioning process and different test temperatures, but are still similar enough to give us confidence that the twofold increase in the number of atypical spots is strongly related to the decrease in temperature.
Similarly, the morphological data in Table~\ref{tab:ratios} for both sets match very closely, giving further assurance in comparison of the atypical features between the two sets. 
We note that the separate group of entries around 3~kV in Fig.~\ref{fig:BD_voltages} is due to an artifact of the conditioning procedure. Following a BD, the applied voltage is dropped to zero and then ramped logarithmically to the previously set target voltage in several steps. It was previously observed that BD might lead to a follow-up event (see, for example, \cite{wuenschStatisticsVacuumBreakdown2017}), which, due to our ramp-up procedure, will be observed at mid-level voltage predefined by the logarithmic ramping.
 
These spots are very characteristic, with a very shallow depth and star-like shape instead of circular ones. The size of the star-like feature is harder to measure, but fitting a circle to the largest dark part of the feature gives values somewhat smaller than the regular one, but not significantly. The corresponding anode sites are also more shallow than the regular ones, but similar in morphology.

H.~Padamsee and J.~Knobloch observed the starbursts on BD sites formed in a superconducting RF cavity made from Nb and Nb electrodes tested under DC~\cite{Padamsee:1998ya}. 
They proved with SEM and Auger spectroscopy that the starbursts were sites in which a very thin film of fluorine contamination was burnt by plasma ignited during the BD event. 
Thus, the starburst was a fingerprint left by a plasma cloud on a thin pre-existing contamination layer. 
The star-like BDs observed in the present work are due to deposited copper and create a significant impact on the surface, leading to complete covering of the machining groves. 
This could be a result of the lower melting temperature of Cu relative to Nb (1085~$^{\circ}$C vs 2477~$^{\circ}$C) combined with the difference in energy available for the evolution of the BD spot. In the current study, we do see a symmetric imprint left of the thin surface layer of contamination/oxidation beyond the rough circumference of the star-like BD.

Therefore, we speculate that the starburst feature is a result of instability in the evolution of the Cu plasma cloud created during the BD process. 
This instability leads to explosive nonuniform deposition of melted Cu on the surface, which, combined with the low temperature of the cathode, leads to the frozen asymmetrical star-like feature. 
The same plasma cloud also leads to surface melting and crater formation.
For electrodes at higher temperatures (room temperature in our study), late-stage melting of the surface within the large circular region and the motion of ejected molten copper from the crater cover the star-like marking, which was formed earlier. 
Thus, in room-temperature experiments, we did not find any traces of the instabilities; however, in cryogenic temperatures, the heat is dissipated much more efficiently due to much higher thermal diffusivity, $10^{-4} \frac{m^2}{s}$ at 300~K and $ 0.2~\frac{m^2}{s}$ at 10~K~\cite{Ekin2006}, preventing the formation of a large melted copper spot and allowing cryogenically cooled copper to rapidly crystallize on the surface, leaving clear traces that can provide a glimpse of the initial stages of plasma evolution.

The difference in morphology between the anode spots and the cathode is quite baffling. According to standard models, these are the result of the expansion of a plasma cloud and are expected to form a similar spot structure on both sides~\cite{sladeVacuumInterrupterTheory2020}.
We note that the breakdown spot for a typical BD event is approximately four times larger than the gap size (taking average spot diameter into account), indicating that, given the gap geometry, the plasma occupies a significant portion of the gap around the emission spot — extending well beyond the expected size of its initial size.
Indeed, the cathode spots show clear melting and reshaping of the surface. 
However, the anodes' spots, demonstrate a more complicated morphology. The machining traces are visible on the innermost, lighter disk of the anode, indicating that no significant melting took place within this region. Around the disk, a rough part, which was clearly melted, covers the surface machining grooves; see Fig.~\ref{fig:Typical_BD}(a).
In addition, while on the cathode side, the spot is surrounded by melted ejecta in the form of jets; these are missing on the anode side. 
Lastly, we find no significant differences between the room temperature and cryogenically conditioned anode spots.

The observation of a central unmelted region on the anode side can be explained if the plasma plume that approaches the anode's surface is diverted radially from the center. 
We suggest that this shielding effect, which is missing from the current VBD description, might result from our system's specific geometry. Still, clarifying this can enhance our overall understanding of late-stage BD evolution.

Assuming that the plasma originated from a spot on the cathode and went through a rapid expansion into vacuum, we would expect it to have a spherically symmetric distribution that follows a cosine distribution with respect to the angle to the surface normal, $\nu$~ \cite{andersAngularlyResolvedMeasurements2002}.
Thus, one would presume to find the maximum of the plasma density, $n$, at the center of the anode spot with a quadratic drop with the distance from the center, $r$~\cite{anders_cathodic_arcs}:

\begin{equation}
n = C\frac{I_{arc}}{r^2} \cos{\nu}
\end{equation}
where $C$ is a constant (for copper $C \approx 10^{13} \frac{1}{Am}$) and $I_{arc}$ is the arc current.
This is clearly not the pattern we are observing on the anode surface.

In most previous studies that measured the angular distributions of the plasma plume after vacuum breakdown, large gap systems were used to allow better observation of plasma evolution, assuming that the gap distance does not have a significant effect on the results (see, for example,~\cite {andersAngularlyResolvedMeasurements2002} and references therein).
Similar considerations led to large gap experiments, with gaps of $>1$~mm for plasma spectroscopy during BD ~\cite{daviesEmissionElectrodeVapor2008, Zhou2021}.
This allows for the observation that the inter-electrode glow started from the cathode.
However, in all of these, the formation of a central shielded region on the anode side was not reported. 
The shielding effect we observe could be a unique result of the current configuration and induced before the plasma had fully formed, in the initial stage of the BD, which starts with localized field emission.
To examine this hypothesis, we estimate the effect of the electrons emitted during this stage on the anode in our geometry, which has a small gap and relatively low applied voltage.

In large gap experiments, the initial beam of field-emitted electrons is not narrow enough to heat and start evaporation from the anode and has too large energy ($>$50~keV) for efficient gas ionization. 
In the smaller gap system we use, the beam is still narrow when reaching the anode, thus depositing a locally higher energy density, and combined with lower electron energies, has a higher probability of gas ionization.
 Assuming a uniformly circular emission area with a radius of $r_e$, and a region where the field is locally enhanced, $r_p$, it can be shown that the beam radius at a distance $d$ from the emitter will be~\cite{sladeVacuumInterrupterTheory2020}:

\begin{equation}
r_a = 2r_e \left( \beta \frac{d}{r_p} \right) ^{1/2}
\end{equation}
For simplicity, we assume that $r_p = r_e = 10$~$nm$ and take the $\beta = 96$, as measured before using field emission, which leads to a high space charge region with a radius of 15.3~$\mu m$ on the anode surface. 
Such a space charge may lead to the formation of a local shielded region, as observed in Fig.~\ref{fig:Typical_BD}(a) and Table~\ref{tab:ratios}).

\section{Conclusions}

The effects of conditioning on electrodes were studied at room and cryogenic temperatures, demonstrating an improvement in field retention capacity at cryoconditioned samples. 
These were accompanied by significant variations in the properties of post-breakdown damage to the surfaces.
Although the microhardness of the surface was modified at breakdown-damage sites, we demonstrate that it is not related to conditioning, as the microhardness is similar between conditioned and unconditioned regions. 

Comparative sub-surface microscopy of areas that were not visibly damaged due to breakdown showed marked differences in the material's structure in correlation with exposure to high field. 
Though no damage was seen from the top, as measurements were done between BD spots,  clear changes to the dislocation structure were observed compared to the reference regions on the edge of the electrode. 
These changes, which we suggest are related to conditioning, regardless of temperature, resulted in a visible decrease and not an increase in the number of dislocation walls.
We speculate that this is due to dislocation activity induced by the sub-yield stresses generated by the external field. Notably, this effect was stronger in the cold-conditioned samples. 
This observation is consistent with breakdown initiation being linked to sub-surface plastic activity mediated by dislocations. 
We speculate that the ultimate breakdown strength of copper is achieved after a period of surface conditioning, during which the near-surface structure evolves. 
This period is longer when one starts with heat-treated, 'soft' copper and shorter for 'hard' copper (see~\cite{korsback2020}), which is reasonable due to the high availability of defect sources and sinks in such systems. 
The higher field retention ability observed at cryo temperature is consistent with previous observations \cite{Cahill2018a,Jacewicz2020} and theoretical models based on defect and dislocation kinetics \cite{Engelberg2018PRL,engelberg_theory_2019,Nordlund:2012zz}.

These observations indicate the complex nature of the underlying process controlling conditioning, leading to the need for a further systematic study of surface modification and its dependence on applied fields and temperatures. These call for additional methods that allow for in situ monitoring of exposed surfaces, such as surface impedance monitoring ~\cite{comanSituResistivityMeasurement2023}.

%\nocite{*}

\bibliography{ColdCuref.bib}

%\bibliography{references.bib}
\end{document}